\documentclass[paper]{JHEP3}

\usepackage{epsfig}
\usepackage{rotating}
\usepackage{amssymb}
\usepackage{moreverb}
\usepackage{amsmath}
\usepackage{amsbsy}
\usepackage{enumerate}
\usepackage{bbm}
\usepackage{url}
\usepackage{cite}

\newcommand\sss{\mathchoice%
{\displaystyle}%
{\scriptstyle}%
{\scriptscriptstyle}%
{\scriptscriptstyle}%
}

\newcommand\as{\alpha_{\sss\rm S}}
\newcommand\mur{\mu_{\sss\rm R}}
\newcommand\muf{\mu_{\sss\rm F}}
\newcommand\GF{G_{\sss\rm F}}
\newcommand\aOLP{\alpha_{\sss\rm OLP}}
\newcommand\aIR{\alpha_{\sss\rm IR}}

\def\({\left(}
\def\){\right)}
\def\LO{\rm LO}
\def\NLO{\rm NLO}

\title{A proposal for a standard interface between \\
Monte Carlo tools and one-loop programs}

\author{T. Binoth
\\
The University of Edinburgh,
Edinburgh EH9~3JZ,
Scotland, United Kingdom}

\author{F. Boudjema \footnote{Corresponding author: boudjema@lapp.in2p3.fr}\\
LAPTH, Universit\'e de Savoie, CNRS, 9 Chemin de Bellevue BP 110,
74941 Annecy-le-Vieux, France}

\author{G. Dissertori and A. Lazopoulos\\Department of Physics, ETH
  Zurich, CH--8093 Zurich, Switzerland}

\author{A. Denner\\Paul Scherrer Institut, W\"urenlingen und Villigen,
 CH-5232 Villigen PSI, Switzerland}

\author{S. Dittmaier\\Albert-Ludwigs-Universit\"at Freiburg, Physikalisches
    Institut, D-79104 Freiburg, Germany}

\author{R. Frederix, N. Greiner and S. H\"oche\\
University of Zurich, Winterthurerstrasse 190, CH-8057 Zurich, Switzerland}

\author{W. Giele, P. Skands and J. Winter\\FNAL, P.O. Box 500, Batavia, IL
  60510, USA}

\author{T. Gleisberg\\SLAC, Stanford University, Stanford, CA 94309 USA}

\author{J. Archibald, G. Heinrich, F. Krauss and D. Ma\^{\i}tre\\IPPP,
  University of Durham, South Rd, Durham DH1~3LE, United Kingdom}

\author{M. Huber\\
Max-Planck-Institut f\"ur Physik (Werner-Heisenberg-Institut),
D-80805 M\"unchen, Germany}

\author{J. Huston\\Michigan State University, East Lansing, MI 48840, USA}

\author{N. Kauer\\Department of Physics, Royal Holloway, University of
  London, Egham TW20 0EX, UK}

\author{F. Maltoni\\
CP3, Universit\'{e} Catholique de Louvain,
B--1348 Louvain-la-Neuve, Belgium}

\author{C. Oleari\\Universit\`a di Milano-Bicocca, 20126 Milano, Italy}

\author{G. Passarino\\INFN, Sez. di Torino and Dip. Fisica Teorica,
      Universit\`a di Torino,
Torino, Italy}

\author{R. Pittau\\Departamento de F´isica Te´orica y del Cosmos CAFPE, Universidad de Granada, E-18071 Granada, Spain}

\author{S. Pozzorini\\CERN/PH, CH--1211 Geneva 23, Switzerland}

\author{T. Reiter\\Nikhef, Science Park 105, 1098 XG Amsterdam, The
  Netherlands}

\author{S. Schumann\\Institut f\"ur Theoretische Physik, Universit\"at
  Heidelberg,
D-69120, Heidelberg, Germany}

\author{G. Zanderighi\\ Rudolf Peierls Centre for Theoretical Physics, 1
  Keble Road, OX13PN, Oxford, UK}

\abstract{Many highly developed Monte Carlo tools for the evaluation of cross
  sections based on tree matrix elements exist and are used by experimental
  collaborations in high energy physics.  As the evaluation of one-loop
  matrix elements has recently been undergoing enormous progress, the
  combination of one-loop matrix elements with existing Monte Carlo tools is
  on the horizon. This would lead to phenomenological predictions at the
  next-to-leading order level.  This note summarises the discussion of the
  next-to-leading order multi-leg~(NLM) working group on this issue which
  has been taking place during the workshop on Physics at TeV colliders
  at Les Houches, France, in June 2009.  The result is a proposal for a
  standard interface between Monte Carlo tools and one-loop matrix element
  programs.

\vspace{0.7cm}
{\em

Dedicated to the memory of, and in tribute to, Thomas Binoth, who
led the effort to develop this proposal for Les Houches 2009.
Thomas led the discussions, set up the subgroups, collected the
contributions, and wrote and edited this paper. He made a promise
that the paper would be on the arXiv the first week of January,
and we are faithfully fulfilling his promise. In his honour, we
would like to call this the Binoth Les Houches Accord.

}

}


\begin{document}

\section{Introduction}
\label{intro}
Monte Carlo event generators are of major importance in high energy physics
as tools to simulate scattering processes defined within the Standard Model
or one of its extensions.  Probability distributions of events are determined
from matrix elements which are defined in perturbation theory. At present
there exists a plethora of highly developed Monte Carlo~(MC) event generators
which are based on tree matrix elements and which use sophisticated methods
for matrix element generation and phase space
integration~\cite{Alwall:2007st, Cafarella:2007pc, Dobbs:2004qw,
Gleisberg:2008fv, Kilian:2007gr, Krauss:2001iv, Maltoni:2002qb,
Mangano:2002ea, Papadopoulos:2006mh, Pukhov:1999gg, Stelzer:1994ta, calcHEP,
GRACE, Fujimoto:2002sj}.  In combination with parton showers and
hadronisation models, realistic event distributions can be
obtained~\cite{Bahr:2008pv, Gleisberg:2008ta, Sjostrand:2006za,
Sjostrand:2007gs, Corcella:2000bw, Corcella:2002jc}.  In predicting absolute
rates, these leading order calculations are plagued by uncertainties due to
the missing higher order terms. Note that $N$-jet
observables scale like $\as^N(\mur)$ and are thus very sensitive to the
renormalisation scale~($\mur$) choice which can not be inferred from first
principles. In the case of electroweak corrections, higher order terms can
become sizable at high energies, i.e. of the order of ten percent, due to
large electroweak Sudakov logarithms.  For precise predictions and
simulations of collider data, it would be highly desirable to take such
corrections into account.

Together with real-emission corrections, virtual corrections contribute to
NLO calculations too, so that one is faced with the problem of calculating
one-loop Feynman diagrams The evaluation of this family of diagrams has seen
major progress in the last few years. Especially in the last two years, many
processes of the Les Houches experimenters wishlist have been
evaluated~\cite{Berger:2009zg, Berger:2009ep, KeithEllis:2009bu,
Melnikov:2009wh, Campbell:2006xx, Campbell:2007ev, Ciccolini:2007jr,
Bredenstein:2009aj, Dittmaier:2009un, Dittmaier:2007wz, Dittmaier:2007th,
Andersen:2007mp, Binoth:2008kt, Binoth:2009rv, Binoth:2009wk,
Campanario:2008yg, Campanario:2009um, Jager:2009xx, Bozzi:2009ig,
Bevilacqua:2009zn, Arnold:2008rz}.  Until recently, the bottleneck has been
the numerically stable and efficient evaluation of these one-loop matrix
elements.  Unitarity-based methods, but also improvements of
Feynman-diagram-based methods, have now matured to the point where the
evaluation of the one-loop matrix elements is no longer the stumbling block.
As stated above, the evaluation of tree-matrix elements is well under control
and the corresponding MC tools can be used for the evaluation of the leading
order~(LO) and next-to-leading order~(NLO) real emission contributions in
full NLO computations. Together with methods to treat the inherent infra-red
singularities, everything is in place to combine tools of different
communities and positive synergy effects can be expected from such an
initiative.

Various groups, using different methods, are meanwhile able to make NLO
predictions for partonic processes with up to four final-state
particles~\cite{Berger:2009zg, KeithEllis:2009bu, Bredenstein:2009aj,
Binoth:2009rv, Bevilacqua:2009zn}.
Several groups aim for an automated evaluation of one-loop matrix
elements~\cite{Hahn:1998yk, Kurihara:2002ne, Belanger:2003sd, Ellis:2007br,
Ossola:2007ax, Binoth:2008uq, Berger:2008sj, Lazopoulos:2008ex,
Winter:2009kd} or provide program packages for certain classes of
processes. Thus a good coverage of next-to-leading order processes relevant
for the LHC is in reach.

The incorporation of one-loop matrix element information into existing
tree-level Monte Carlo tools is feasible due to the inherent modularity of
higher order calculations, as will be discussed below.  Note that such a
synthesis would speed up the production of NLO results considerably because
the loop calculator would be dispensed from dealing with the tree-like parts
of the computation and phase-space integration issues. Of course, numerical
stability of the one-loop matrix element, at least in a certain region of
phase space (to be specified by the provider), has to be guaranteed.  For the
experimentalist, such a coherent approach would allow to use familiar tools,
instead of dealing with many different stand-alone computer codes.

The members of the working group on next-to-leading order multi-leg processes
have extensively and constructively discussed this issue during the Les
Houches workshop in June 2009. This document summarises the discussion of the
working group and provides a proposal for a standardised interface between
Monte Carlo tools and one-loop programs.

A short outline of the structure of one-loop computations is provided in
section~\ref{sec:modstruc} to set the stage for the proposal of the interface
in section~\ref{sec:interface}. Section~\ref{sec:OLP_interface} contains
details on how to implement such an interface and in section~\ref{sec:EWcase}
an extension to electroweak corrections is discussed.
The article closes with a summary and gives an outlook on further
extensions of the proposed Les Houches Accord.

\section{Modular structure of one-loop computations}
\label{sec:modstruc}
Higher order corrections to scattering processes require the evaluation of
virtual and real emission corrections for a given process. These corrections
have to be provided for all contributing partonic cross sections.  In
hadronic collisions the hadronic cross section is given by
\begin{eqnarray}
 \sigma_{\sss had}(p_1,p_2) &=& \sum\limits_{a,b}
\int\limits dx_1 \, f_{a/H_1}\!\(x_1,\muf^2\)
\int\limits dx_2 \, f_{b/H_2}\!\(x_2,\muf^2\)  \nonumber\\
&&  \times
\Big[ d\sigma^{\LO}_{ab}\!\(x_1p_1,x_2 p_2;\mur^2\) +
d\sigma^{\NLO}_{ab}\!\(x_1p_1,x_2 p_2;\mur^2,\muf^2\)  \Bigr] ,
\end{eqnarray}
where $a,b$ label the partonic subprocesses.  Note that, in the case of
electroweak corrections, photons have to be allowed in the initial state at
next-to-leading order in $\alpha$.  The partonic contributions consist of the
Born term~($B$), real~($R$) and virtual~($V$) corrections and collinear
counterterms~($C$).  For observables with $m$ particles in the final state at
LO, the real-emission term contains one extra particle in the final state.
\begin{eqnarray}
\sigma_{ab}^{\LO} &=& \int_m d\sigma_{ab}^B\,, \\
\sigma_{ab}^{\NLO} &=& \int_{m+1} d\sigma_{ab}^R + \int_m d\sigma_{ab}^V +
\int_m d\sigma_{ab}^C\!\(\muf^2, {\rm F.S.}\).
\end{eqnarray}
The collinear counterterm, $d\sigma_{ab}^C(\muf^2, {\rm F.S.})$, is induced
through a perturbative reparametrisation of the parton distribution functions
and cancels the collinear initial state singularities of the real emission
contribution. It is actually a convolution of the Born term with a splitting
function which depends on a factorisation scheme~(F.S.) and introduces a
factorisation scale $(\muf)$ dependence.
In the case of fragmentation processes, a similar treatment for final-state
hadrons and/or photons apply and the fragmentation scheme has to be
specified.

The Born, real emission and collinear terms are defined by $2\to m $ and $2
\to m+1$ tree amplitudes, which can be efficiently evaluated with existing
matrix element generators.  The virtual contribution is the integral over the
$m$-particle final state of the interference term, ${\cal I}(\{k_j\})$,
between the Born and the one-loop amplitude
\begin{eqnarray}
d\sigma_{ab}^V &=& d\,{\textrm{LIPS}}(\{k_j\})\;  \mathcal{I}(\{k_j\})\,.
\end{eqnarray}
Here $\{k_j\}$ with $j=1,\dots,m+2$ indicate the partonic momenta.  The
interference term is defined by the tree- and one-loop contribution of the
given process. These parts share a common colour basis, $| c \rangle$, which
defines the corresponding colour correlation matrix, $\langle c | c'\rangle$.
\begin{equation}
\mathcal{I}(\{k_j\}) = \sum\limits_{h,c,c'} \left( \;\mathcal{A}^{{\LO}
  \dagger}_{h,c} \langle c | c' \rangle \mathcal{A}^{{\NLO}, V}_{h,c'} +
\mathcal{A}^{{\NLO}, V \dagger}_{h,c} \langle c | c' \rangle
\mathcal{A}^{\LO}_{h,c'}\; \right) \,.
\end{equation}
The indices $h$ and $c$ enumerate helicity and colour degrees of freedom.
Different representations of the underlying amplitudes are possible. For
example, specifying a helicity and colour basis may allow for Monte Carlo
sampling over different components of the cross section contributions. On the
other hand, the summation over colour and helicity leads to a simple,
basis-independent object.  In some cases it might be sufficient to provide
approximations for the loop amplitude, e.g.\ the leading colour term(s).  The
one-loop amplitude contains ultra-violet~(UV) and infra-red~(IR) divergences.
Dimensional regularisation converts these divergences into poles in
$\epsilon$ where, typically, the convention $d=4-2\epsilon$ is adopted for
the space-time dimensionality.  Applying a UV renormalisation scheme
removes the UV poles from the loop amplitude and one is left with IR
singularities. Those are very often treated by using dimensional
regularisation as well, which converts them into IR $\epsilon$ poles but, of
course, other methods are applicable too. For example, mass regulators are
often used in QED and electroweak computations, as will be discussed below.
Any IR regularisation scheme~(R.S.) defines uniquely the finite remainders
(once an overall constant factor is factorized in front).  Commonly used
schemes differ by the dimensional treatment of internal and external
particles.
Standard examples are conventional dimensional regularisation~({\tt
  CDR}), dimensional reduction~({\tt DRED}) and the
't~Hooft--Veltman scheme~({\tt tHV})~\cite{Catani:1996pk}.  Schemes where the
external particles are 4-dimensional are preferable from the point of view of
helicity amplitudes. {\tt CDR} and {\tt tHV} have the same dimensional
treatment concerning the soft/collinear particles.  For a discussion of
different schemes, see e.g.~\cite{Kunszt:1993sd, Catani:1996pk, Smith:2004ck,
  Signer:2008va}.  In the context of electroweak one-loop corrections,
different computational schemes are used to take into account generic loop
contributions and the treatment of unstable particles. This will be
extensively discussed in section~\ref{sec:EWcase}.

Any standardised treatment of these issues should allow for maximal
flexibility to choose an adequate method and scheme for a computation.  The
only general assumption we adopt in our proposal is that the UV
renormalisation is done internally by the provider of the one-loop
contribution and that the IR singularities are treated uniformly such that
interfacing with simple information passing is possible.

Due to its wide acceptance we will describe here the use of dimensional
regularisation for IR divergences in more detail.  In
section~\ref{sec:EWcase} on electroweak~(EW) corrections, we will discuss how
regulator masses can be treated with the same general interface.  If
dimensional regularisation is applied, one is left with IR poles only, after
the UV renormalisation.  The general structure of the virtual correction is
thus given as
\begin{equation}
\label{LaurentSeries}
\mathcal{I}\!\(\{k_j\},{\rm R.S.},\mur^2,\as(\mur^2),\alpha,\dots\) =
C(\epsilon) \left(\frac{A_2}{\epsilon^2} + \frac{A_1}{\epsilon}+ A_0 \right) .
\end{equation}
The Laurent coefficients $A_j$ are real-valued. Apart from the regularisation
scheme and the renormalisation scale $\mur$, they depend of course on all
relevant couplings and masses of the underlying model which is indicated by
the dots in the argument list of eq.~(\ref{LaurentSeries}).  A choice for the
overall constant which is used by several authors is
\begin{eqnarray}
C(\epsilon) = \frac{(4\pi)^\epsilon}{\Gamma(1-\epsilon)}
\(\frac{\mu^2}{\mur^2}\)^\epsilon = (4\pi)^\epsilon
\frac{\Gamma(1+\epsilon)\Gamma(1-\epsilon)^2}{\Gamma(1-2 \epsilon)}
\(\frac{\mu^2}{\mur^2}\)^\epsilon + \mathcal{O}\(\epsilon^3\) \, .
\end{eqnarray}
where $\mu$ is the dimensional regularisation scale and $\mur$ is the
renormalisation scale.  These scales are often identifies, i.e.~$\mu=\mur$.
Other choices will not lead to any problem as long as the information is
transparent.  The remaining infrared poles cancel when the virtual part is
combined with the real emission corrections after integration over
soft/collinear phase-space regions and with the collinear counterterms.  In
actual calculations, subtraction methods are applied which define local
subtraction terms for the real-emission amplitude and add back the integrated
version, such that IR divergences cancel with the $m$ and $m+1$ particle
contributions separately.  Two prominent examples are the widely used
Catani--Seymour dipole subtraction method~\cite{Catani:1996vz, Phaf:2001gc,
Catani:2002hc} and the FKS subtraction method~\cite{Frixione:1995ms,
Frixione:1997np}, which have been implemented in computer programs by several
groups~\cite{Seymour:2008mu, Gleisberg:2007md, Frederix:2008hu,
Czakon:2009ss, Hasegawa:2009tx, Frederix:2009yq}.

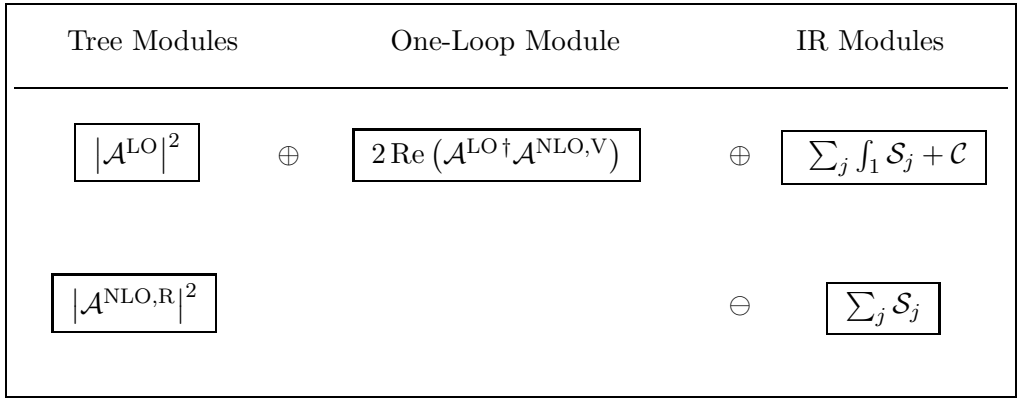
\begin{figure}[htb]
\setlength{\unitlength}{1mm}
\begin{center}
\fbox{\begin{picture}(132,50)
\put(7,45){Tree Modules}
\put(50,45){One-Loop Module}
\put(104,45){IR Modules}
\put(0,40){\line(1,0){132}}
\put(8,30){\fbox{ $\left|\mathcal{A}^{\LO}\right|^2$ }}
\put(35,30){$\oplus$}
\put(45,30){\fbox{ $2 \, {\rm Re}\( \mathcal{A}^{\LO\, \dagger}
    \mathcal{A}^{\NLO, V} \)$ }}
\put(95,30){$\oplus$}
\put(102,30){\fbox{ $\; \sum_j \int_1 \mathcal{S}_j + {\cal C} $ }}
\put(5,10){\fbox{ $\left |\mathcal{A}^{\NLO, R}\right|^2$ }}
\put(95,10){$\ominus$}
\put(108,10){\fbox{ $\sum_j \mathcal{S}_j$ }}
\end{picture}}
\end{center}
\caption{Modular structure of next-to-leading order computations for partonic
  processes.  All structures related to tree amplitudes can be evaluated
  using LO MC tools. The one-loop module contains the UV renormalised
  interference term. The treatment of IR subtraction should be kept separate
  to allow for flexibility. The IR modules contain subtraction terms, ${\cal
    S}_j$, for the real-emission part and their integrated variants which
  compensate IR divergences in the One-Loop Module.  In case of collinear
  initial-state divergences, collinear subtraction terms, ${\cal C}$, have to
  be provided too.  Subsequently, the contributions in each horizontal line are
  independently finite after summation.}
\label{modnlo}
\end{figure}

The general modular structure of a one-loop computation is summarised in
Fig.~\ref{modnlo}.

There are cases where a partonic process is not allowed at tree level but is
initiated at the one-loop level only. In these cases, Fig.~\ref{modnlo}
contains only a one-loop module, the squared one-loop amplitude
$|\mathcal{A}^{\LO}_{{\rm One\mbox{-}Loop}}|^2$, instead of the
interference term.

After this summary of the general structure of one-loop computations, we are
in the position to discuss our proposal for an interface between Monte Carlo
event generators, based on tree matrix elements, and generators of one-loop
amplitudes.

\section{A computational model for an interface}
\label{sec:interface}
The goal of the interface is to facilitate the transfer of information
between one-loop programs, abbreviated as OLP in the following, and programs
which provide tree amplitude information and incorporate methods to perform the
integration over the phase space.  We call the latter simply Monte Carlo tool
(MC) in the following.  It has to be stressed that any agreement on a common
interface should avoid to constrain a provider of an OLP. The idea is to
provide a platform to exchange information between an OLP and a MC in a
compatible way.  The information which has to be passed is already defined in
section~\ref{sec:modstruc}.  Note that the MC and the OLP may work in very
different ways. Some programs might either evaluate one-loop amplitudes or
produce amplitude code on the fly as requested by a user, while others may
consist of a library of hard coded matrix elements for a certain number of
processes.

The proposed computational model to fix the interaction between an OLP and a
MC works in two phases: the {\tt initialisation} and the {\tt run-time}
phase.

\begin{description}
\item[{\tt Initialisation}:] During this phase the MC communicates basic
  information to the OLP.  This means it will ask for the availability of
  partonic one-loop sub-processes, set up the input parameters and fix
  options that are offered by the OLP.
\item[{\tt Run-time}:] During this phase, the MC queries the OLP for the
  value of phase-space dependent one-loop contributions and additional
  information.  The latter should also contain a return value for the squared
  LO amplitude which is useful for checking purposes and gives flexibility in
  the application of different renormalisation/factorisation schemes.  The
  most relevant information is the finite part of the interference term.  It
  may be split into several contributions to allow for more efficient Monte
  Carlo integration sampling.
\end{description}

We propose to pass the in/output information in the {\tt initialisation}
phase through a file and during the {\tt run-time} phase by function
calls. This will be explained in more detail below.  We propose to pass model
parameters as a Les Houches Accord file with its specific
format~\cite{Skands:2003cj, Allanach:2008qq} since it is widely used in the
community.  Schematically the input/output information is collected in
Tab.~\ref{inoutcontrol}.
\begin{table}[htb]
\begin{center}
\begin{tabular}{|l|l|l|}
\hline
OLP & Input & Output \\[1mm]
\hline \hline
 & Model parameters: &  \\[1.5mm]
 & $\alpha(0),\, \as(M_Z),\dots$, $m_t,\,  m_b, \dots$, CKM values &   {\tt
  confirm values} \\\cline{2-3}
 & Schemes: & \\[1.5mm]
 & UV-renormalisation / IR-factorisation   & {\tt confirm schemes} \\ \cline{2-3}
 & Operational information: & \\[1.5mm]
\rule{1mm}{0pt}\ \ \
\begin{rotate}{90} \ \  {\tt Initialisation} \end{rotate}\rule{1mm}{0pt} &
colour/helicity treatment,
approximations, etc. & {\tt confirm options} \\
\hline\hline
 & Events: &  \\[1.5mm]
 & $( E,p_{x},p_{y} ,p_{z},M )_{j=1,\dots,m+2},\, \mu,\, \as(\mur)$ &
 $\(A_2,A_1,A_0,|\textrm{Born}|^2\)$ \\[1.5mm]
\rule{1mm}{0pt} \ \
\begin{rotate}{90} {\tt Run-time}\ \  \end{rotate} \rule{1mm}{0pt} & & {\tt
  optional information} \\
\hline
\end{tabular}
\end{center}
\caption{\label{inoutcontrol}The basic input/output information of a one-loop
  program. In the initialisation phase, communication is preferably achieved
  through files, whereas, in the run-time phase, the information is passed by
  function calls.}
\end{table}

A few comments are in order.
\begin{itemize}
\item Input parameters should generally be provided by the MC during the
  initialisation phase.  We recommend that for electroweak calculations the
  fine-structure constant is passed by the MC as $\alpha(0)$. EW computations
  might entail modifications here as discussed in section~\ref{sec:EWcase}.
\item The IR regularisation schemes can be labelled by keywords like {\tt
  CDR}, {\tt DRED}, {\tt tHV}, etc.
\item Any approximation which is used in the OLP has to be transparently
  defined and implemented such that it can be flagged during
  initialisation. Examples are the use of massless approximations for certain
  quarks or a diagonal unit CKM matrix.
\item The minimal one-loop information to be returned by the OLP is the
  colour- and helicity-summed interference term (at least in some
  approximation).  Additional information can be returned too. It is the
  responsibility of the OLP author to define the different additional return
  values/options of the code in its documentation.
\item Further keywords maybe used for the treatment of massive or unstable
  particles or for EW schemes. See section~\ref{sec:EWcase} for further
  details.
\end{itemize}

\section{Proposal for the implementation of the MC/OLP interface}
\label{sec:OLP_interface}
This section outlines a proposal for the implementation of the interface
model presented above.  Of course any proposal should allow for easy
implementation in all common programming languages.

In the easiest case of merging two programs written in a compatible language,
the connection between the caller (the MC) and the receiver (the OLP) can be
achieved by linking the program together.  In more difficult cases, with
mixed languages or special environments, a bridge program, or transmission
protocol, will have to be put in place. This is, however, not the topic or
concern of this proposal.  We only want to ensure that the in/output values
are defined in a mutually compatible way.

In the initialisation phase, communication between the MC and the OLP will
happen such that availability or generation of amplitude information of
specified processes, including necessary settings and options, is
communicated.  After successful initialisation, the codes are prepared to
pass actual amplitude information for cross section evaluations in the {\tt
  run-time} phase.  We propose to pass loop-amplitude information by function
calls located in the MC and executed by the OLP.  How the function is
actually implemented is left to the OLP but we propose to use the generic
name {\tt OLP\_EvalSubProcess} and to fix the arguments and return values
following Tab.~\ref{inoutcontrol}, as specified below.

In detail we propose the following set-up for the implementation of the two
interaction phases of the MC with the OLP.

\subsection{The initialisation phase}
In this phase, the MC creates a file that contains the information about the
subprocesses it will need to perform the computation. As subprocess we
understand either a partonic subprocess or a component thereof. To introduce
such components is useful if, for example, detailed colour or helicity
information is provided by the OLP such that sampling methods can be
applied. The subprocesses are represented by integer numbers which are
allocated in the initialisation phase.  We will call the generated file the
{\tt order}.  The choice to transfer the set-up parameters through files
allows to invoke the initialisation of the MC and the OLP, without having to
link the two programs together.  The order file is then read by the OLP and
it creates a new file with the agreement about how to link the MC and OLP at
run-time. We will call this file the {\tt contract}.

\subsubsection{The {\tt order} file}
The order file contains the information needed for the OLP to declare whether
or not it can provide the subprocess the MC asks for. It should be allowed,
for example, to ask for colour- and helicity-summed amplitudes.  The first
part of the file contains some general settings as defined in the {\tt
  initialisation} part of Tab.~\ref{inoutcontrol}.

The settings that an OLP has to provide concern the colour and helicity
information and the treatment of divergences.
\begin{description}
\item[{\tt MatrixElementSquareType}:] the type of the returned amplitude
  information.  This flag can be used to distinguish colour~($C$) and
  helicity~($H$) treatment. Possible values could be {\tt CHsummed}, {\tt
    Csummed}, {\tt Hsummed}, {\tt NOTsummed}.  Note that helicity/colour
  basis information has to be provided if summations are not
  performed. Similar flags for colour/helicity averaged expressions like {\tt
    CHaveraged}, etc. might also be used.
\item[{\tt CorrectionType}:] the type of higher order correction should be
  specified. Standard keywords are {\tt QCD}, {\tt EW} or {\tt QED}.
\item[{\tt IRregularisation}:] the IR regularisation scheme used.  Common
  choices for QCD are {\tt CDR}, {\tt DRED}, {\tt tHV}, as explained in the
  text. The OLP provider might define his/her scheme of choice, as long as it
  is clearly documented.
\end{description}

Further settings which are related to EW processes are discussed in
section~\ref{sec:EWcase}.  In the context of EW computations, mass regulators
({\tt MassReg}) or mixed approaches are common to handle IR divergences.
Note that it is assumed that the OLP produces UV renormalised output such
that the MC user should not have to deal with that.  One has to keep in mind,
though, that wave-function renormalisation constants do contain IR
contributions, when external particles are put on-shell, and that IR
subtraction schemes, in general, make assumptions on the treatment of these
wave-function contributions.  This fact is disguised by the use of
dimensional regularisation for UV and IR divergences.  Thus it might be
useful to allow for flags which indicate the used prescription.
\begin{description}
\item[{\tt MassiveParticleScheme}:] a standard choice is {\tt OnShell}.
\end{description}

As most OLP authors use the cancellation of IR divergences as a check, an IR
subtracted result might also be returned by the OLP in addition to the
unsubtracted output. In this case, a flag and the program documentation
should indicate which IR subtraction scheme has been used.
\begin{description}
\item[{\tt IRsubtractionMethod}:] specifies if an IR subtracted result like
  {\tt DipoleSubtraction},\\ {\tt FKSSubtraction}, {\tt AntennaSubtraction},
  etc.  applied to the one-loop result is available, too. The default should
  be {\tt None}.
\end{description}
Although the IR structure of one-loop amplitudes is universal, we do not
recommend to provide only subtracted information, because details of IR
subtraction methods have to be passed then among the MC, the OLP and the IR
subtraction provider, which spoils the modularity of the whole set-up.

In addition, as pointed out above, other specifications have to be added:
most importantly the model parameters like couplings, masses and widths of
particles.  Depending on the process, one has to define the treatment of
unstable particles, number of active flavours, etc.

Settings that concern parameter, technical and operational information are
\begin{description}
\item[{\tt ModelFile}:] the model file from which parameters have to be
  read. We propose to use files in the Les Houches Accord format, {\tt
    $*$.slh} for passing model
  parameters~\cite{Skands:2003cj,Alwall:2007mw,Allanach:2008qq}.
\item[{\tt OperationMode}:] the operating mode of the OLP. This optional flag
  can be used to specify OLP defined approximations to the one-loop
  contribution, e.g.\ {\tt LeadingColour}, {\tt ReggeLimit}, etc.
\item[{\tt SubdivideSubprocess}:] this flag tells the user if a given process
  is represented in a split form to allow for multi-channel Monte Carlo
  sampling.
\end{description}

The OLP can provide defaults for its settings, so that the {\tt ModelFile}
and {\tt OperationMode} options are not mandatory. In this case, the settings
do not need to be present in the order file. The flag {\tt
  SubdivideSubprocess} is proposed to allow for flexibility in the evaluation
of the one-loop contribution.  It defines whether the OLP returns the
amplitude information cut into parts which have different Monte Carlo
weights, when integrated over phase space.  Sampling over various
contributions like colour, helicity, channels, etc. improves the efficiency
of integrators considerably. It is at the discretion of the OLP to provide
such features.

The second part of the order file contains the list of the partonic
subprocesses the MC needs. In this file, each line is of the form
\begin{center}
\begin{boxedverbatim}
 [PDGcode1]            -> [PDGcode2] ... [PDGcodeM]
 [PDGcode1] [PDGcode2] -> [PDGcode3] ... [PDGcodeM]
\end{boxedverbatim}
\end{center}
The first line describes a $1\to M-1$ decay and the second a $2 \to M-2$
scattering process.  [PDGcodeJ] is the particle data group code of particle
$J$ as defined in~\cite{PDGcodes}.
The initial- and final-state particles do, in general, not fix the powers of
the coupling constants. If this is ambiguous it is useful to specify this
flags like
\begin{description}
\item[{\tt AlphasPower}:] integer which specifies the $\as$ power of the
Born cross section.
\item[{\tt AlphaPower}:] integer which specifies the $\alpha$ power of the
Born cross section.
\end{description}
Some OLP providers leave the evaluation of couplings entirely to the MC
and provide only kinematical information where the coupling parameters
are stripped off. This is useful, as no compatibility issues arise
in this case. We propose to indicate this by an {\tt OperationMode} flag:
{\tt CouplingsStrippedOff} which indicates that overall coupling terms
$\sim \alpha^I \as^J$ are treated as equal to one by the OLP.
It is then the responsibility of the MC to multiply the OLP information by the
correct coupling terms which are defined by {\tt AlphasPower}, {\tt AlphaPower}
and {\tt CorrectionType}.

It is recommended that the OLP adds these default settings to the {\tt
  contract} file.  The OLP can define new settings.  If a setting is
mandatory but has not been provided, the OLP should issue an error message
and write only this error message in the contract file.\\

{\bf Example:} Here is an example of an order file for the partonic $2\to 3$
processes, $gg\to t\bar{t}g$, $q\bar{q}\to t\bar{t}g$ and $q g\to t\bar{t} q$,
needed for the evaluation of $pp\to t\bar{t}$ + jet
\begin{center}
\begin{boxedverbatim}
# example order file

MatrixElementSquareType  CHsummed
IRregularisation         CDR
OperationMode            LeadingColour
ModelFile                ModelInLHFormat.slh
SubdivideSubprocess      yes
AlphasPower              3
CorrectionType           QCD

# g g  -> t tbar g
 21 21 -> 6 -6 21
# u ubar -> t tbar g
  2 -2 -> 6 -6 21
# u g  -> t tbar u
  2 21 -> 6 -6  2
\end{boxedverbatim}
\end{center}
In this example, three partonic subprocesses are listed.  The {\tt
  AlphasPower=3} indicates that the Born process is of order ${\cal
  O}(\as^3)$, which implies that only QCD interactions are
considered. Otherwise $Z$-boson/photon exchange would have been allowed in
$q\bar{q}\to t\bar{t}g$.  The {\tt CorrectionType} indicates that the MC asks
for the virtual ${\cal O}(\as^4)$ contribution to the partonic processes.

The flag {\tt SubdivideSubprocess} indicates that the MC asks if a split
representation is provided.  The fact that not all flags occur in this file
indicates that the MC user has read the OLP instructions and knows the
default. As this is of course the exception, preferably, all flags should be
present, together with a corresponding flag.  Lines starting with a $\#$
character are treated as comments.  Comments are encouraged to increase the
human-readability of the file.

\subsubsection{The {\tt contract} file}
In the initialisation phase, the OLP reads the order file and creates its
``answer'', in the form of a contract file. The contract file is basically a
copy of the order file, but with information from the OLP appended at the end
of each (non content-free) line. We propose the following format.

The OLP answer is separated from the MC request by a ``$|$'' character. These
answers are explained in the following paragraphs.  For settings, the answer
should be appended as follows: {\tt OK}, if the setting is supported and the
supplied arguments are correct. Errors might be indicated as follows: {\tt
  Error: unknown option} if the option/setting is not recognised; {\tt Error:
  unknown flag} if the option is supported but the keyword/flag is not
recognised/valid.  If a {\tt ModelFile} is not found the error message is:
{\tt Error: file not found}.

For the requested subprocesses, the answer should be the number of separate
contributions that the OLP can compute, which add up to the required
subprocess, followed by the integer labels which identify each of these
contributions. This allows the multi-channel integrator (if present) to take
advantage of importance difference between the provided contributions. The
integer label will be used at run-time to identify the subprocesses. An OLP
which provides this kind of separation might also provide a single subprocess
that is the sum of all contributions. Whether or not to return the
contributions separately could be set using the setting {\tt
  SubdivideSubprocess}.

Here is an example for a contract file where all options and settings are
available and confirmed by the OLP (\&\& indicates line continuation):
\begin{center}
\begin{boxedverbatim}
# example contract file
# contract produced by OLP, OLP authors, citation policy

MatrixElementSquareType CHsummed                | OK
IRregularisation        CDR                    | OK
OperationMode           LeadingColour           | OK
ModelFile               ModelFileInLHFormat.slh | OK
SubdivideSubprocess     yes                     | OK
CorrectionType          QCD                     | OK

# g g -> t tbar g
21 21 -> 6 -6 21       | 2 13 35 # 2 channels: cut-constructable,&
                                        & rational part
# u ubar -> t tbar g
 2 -2 -> 6 -6 21       | 1 29
# u g -> t tbar u
 2 21 -> 6 -6  2       | 3 8 23 57 # 3 channels: leading,&
                         & subleading, subsubleading colour
\end{boxedverbatim}
\end{center}

If a partonic process is split into several contributions, a description of
these different subprocesses is recommended, as indicated in this example.

It would be very useful if the OLP could provide some help in case of
failure, e.g.\ listing the parameters which are accepted.  All default
parameters, which were not specified in the order file, but are relevant for
the evaluation, should be written by the OLP in the contract file.

Here is an example for a contract file where options and settings are not
available:
\begin{center}
\begin{boxedverbatim}
# example contract file
# contract produced by OLP, OLP authors, citation policy

MatrixElementSquareType CHsummed | Error: unsupported flag
# CHaveraged is supported
IRregularisation        DRED     | Error: unsupported flag
# CDR, tHV are supported
OperationMode      LeadingColour | Error: unsupported flag
# see OLP Documentation
ModelFile     FavouriteModel.slh | Error: file not found
# Modelfile is called: SM.slh
SubdivideSubprocess yes          | Error: unsupported flag
# no is supported
CorrectionType           EW      | Error: unsupported flag
# QCD is supported
MyWayOfDoingThings true          | Error: unknown option

# g g -> t tbar g
21 21 -> 6 -6 21    |   Error: massive quarks not supported
# u ubar -> t tbar g
 2 -2 -> 6 -6 21    |   Error: process not available
# u g -> t tbar u
 2 21 ->> 6 -6  2   |   Error: check syntax

\end{boxedverbatim}
\end{center}
It would be useful that all assumptions and settings which are made and are
available in an OLP are added as comments and/or as warning/error
messages. The latter might be appended in the response line.  The detailed
standardisation of error/warning handling is beyond the scope of this
document. A minimal strategy for the OLP is to use only {\tt OK}, {\tt ERROR}
as identifiers and to encourage the MC user to read the OLP documentation.
Preferably, MC users should be able to correct their order file settings from
the error/warning information and the comments provided in the contract file.

\subsection{The run-time phase}
To start the run-time phase, the OLP needs to be reminded of the contract it
committed to. This is especially the case for programs which dynamically
generate their evaluation structure as opposed to libraries of generated
code. To initialise the OLP for the run, we propose to use a function {\tt
  OLP\_Start} which has to be called by the MC.  After this, the MC can start
querying the OLP for the subprocesses. This is done using the function {\tt
  OLP\_EvalSubProcess}. In more detail these two functions are defined by:

\subsubsection*{{\tt OLP\_Start}}
The function {\tt OLP\_Start} should receive a character string
and an integer as arguments, {\tt OLP\_Start(char*,\&int)}. The
character string indicates the name of the contract file and the
integer is set to 1 by the function call in case of success, or to
something else in case of failure. Confirmation messages and/or an
error message like {\tt Error: can not handle contract file} might
be considered.

\subsubsection*{{\tt OLP\_EvalSubProcess}}

The parameters to be passed to the {\tt OLP\_EvalSubProcess} function are (in
this order):
\begin{itemize}
    \item the integer label of the subprocess
    \item an array containing the components of the momenta. The momenta are
      placed in a one dimensional array. We propose to use physical
      scattering kinematics which differentiates between in and outgoing
      momenta, so that $k_1+k_2=k_3+\dots +k_m$.  For each particle, the
      kinematics is specified by a 5-tuple: $(E_j,k_j^x,k_j^y,k_j^z,M_j)$.  A
      full event is specified by an array of $5*m$ double precision numbers
      filled with the $m$ 5-tuples ordered by the particle enumeration.
%
    \item the renormalisation scale, $\mur$, as a double precision number, or
      an array of scales, if different scales need to be passed.
    \item the strong coupling\footnote{Note that $\as(\mur)=1$ can be used to
      indicate that the {\tt MC} multiplies the returned values with the
      adequate coupling constants.}  $\as(\mur)$, or an array of phase-space
      dependent couplings and/or variables.
    \item the array where the computed results are returned.
\end{itemize}

We propose that all dimensional objects use the {\tt GeV} as
standard unit, with the implicit assumption that only the energy
unit has to be defined, i.e.\ the standard choice $\hbar=c=1$ is
understood. The returned array is expected to contain at least
four real-valued double precision numbers
\begin{center}
\begin{boxedverbatim}
PoleCoeff2, PoleCoeff1, PoleCoeff0, BornSquare
\end{boxedverbatim}
\end{center}
which correspond to the colour- and helicity-summed/averaged terms $A_2$,
$A_1$, $A_0$, $|\textrm{Born}|^2$, as defined in eq.~(\ref{LaurentSeries}),
in Tab.~\ref{inoutcontrol} and in the settings of the contract file.  If the
OLP provider wants to hand over more colour and/or helicity information,
these variables can be promoted to arrays of double precision numbers which
contain the corresponding information.  Care has to be taken, that colour and
helicity conventions match.

It is recommended to use the OLP value(s) for $|\textrm{Born}|^2$ to perform
a consistency check between the OLP and MC evaluation of this quantity, as
the compatibility of settings, parameters, and the colour and helicity basis
is efficiently tested in this way.  The OLP provider might want to hand over
additional information, e.g.\ arrays containing other possible decompositions
of the result.  All these features can also be used for importance sampling
over different expressions of the subprocesses.  The length and the content
of the returned array is to be documented by the OLP.

As most OLP authors use the cancellation of IR poles as a check of their
calculation, a value for the IR subtracted interference term, $A_0^{\rm
  sub}$, might be passed in addition.  A flag and/or the program
documentation should indicate which IR subtraction scheme has been used.




\section{Extending the interface to include EW OLP}
\label{sec:EWcase}
This section outlines a proposal for an extension of the interface model to
include NLO EW corrections.  More communication between the MC and the OLP is
necessary because renormalisation and the treatment of unstable particles are
more complicated than in the QCD case.  The role of input parameters in
electroweak computations, in relation with a computational scheme, has to be
carefully understood when combining output from various sources.
This implies that the exchange of information between an OLP and a MC is, by
far, more complex in case EW NLO corrections are included and necessarily
constrain both contractors. For instance, knowledge of the Born amplitude
alone is not enough to construct the divergent parts of the real corrections
(provided by the MC) fully consistent with the virtual ones (provided by the
OLP).  In order to guarantee cancellation of soft and/or collinear IR
divergences and proper re-weighting of all real NLO corrections, more
information has to be provided.

In the following we describe the minimal amount of information that should be
exchanged during negotiation, taking into account that the interface should
not constrain the OLP in a substantial way.


\subsection{Treatment of resonances}
The calculation of EW corrections requires to fix a scheme for the treatment
of resonances. The presently preferred scheme is the complex-mass scheme
which was introduced at tree level in Ref.~\cite{Denner:1999gp} and extended
to one loop in Ref.~\cite{Denner:2005fg}.  It is gauge invariant and
straightforward to implement in the sense that it follows the usual
perturbative calculus of local quantum field theories, i.e. Feynman rules,
counterterms and so on.  It relies, however, on the use of complex masses and
couplings also in the LO matrix element. To efficiently interrelate MC tools
with EW OLPs, we strongly suggest that LO MCs support this
possibility. Nevertheless, we allow for other schemes, like the naive
fixed-width scheme and the pole scheme (see, e.g., Refs.~\cite{Stuart:1991xk,
  Aeppli:1993cb, Aeppli:1993rs, Veltman:1992tm}).  Note, however, that in
this case more information has to be exchanged between the OLP and the LO MC,
like the used gauge for the fixed-width scheme\footnote{Strictly speaking, the
  fixed-width scheme does not respect gauge invariance but for practical
  purposes, the gauge dependence often turns out to be numerically
  insignificant. Note that additional information on the gauge choice has to
  be provided.} or information on the included diagrams in case of the pole
scheme.  This in general requires that the user studies the description
(e.g.\ documented in a README file) of the OLP.

The basic communication happens via the flag {\tt ResonanceTreatment}.
\begin{description}
\item[{\tt ResonanceTreatment}:] this flag defines the treatment of
  resonances.  Standard values are {\tt ComplexMassScheme}, {\tt
    FixedWidthScheme}, {\tt PoleScheme} or an OLP-defined variant thereof.
\end{description}
If the OLP does not support the requested scheme, it should return a list of
supported schemes in the {\tt contract} file.


\subsection{Electroweak regularisation scheme}
In EW computations, the use of mass regulators for collinear and/or soft
divergences are common. The flag {\tt IRregularisation} which is defined
above should be used to communicate with the OLP.  A prominent approach is to
use fermion masses to regulate collinear divergences but use dimensional
regularisation for the soft divergences. Allowing for several keywords for
the IR regularisation flag, all cases can be covered.
The MC supporting {\tt MassReg} has to provide the information of which
masses are to be considered small, and specific details on how regulator
masses have to be applied in the various parts of the full computation.

The entry in the order file may look like:
\begin{center}
\begin{boxedverbatim}
IRregularisation  MassRegColDiv, DimRegSoftDiv
IRregulatorMasses MD, MU, MS, ME
\end{boxedverbatim}
\end{center}
or
\begin{center}
\begin{boxedverbatim}
IRregularisation  MassReg
IRregulatorMasses MD, MU, MS, MC, MB, ME, MMU, MTAU, MPHOTON
\end{boxedverbatim}
\end{center}
%

Mass regularisation is, for instance, relevant for processes with isolated
muons in the final state\footnote{ Note that dimensional regularisation for
  QED-like soft divergences is fully equivalent to the regularisation by an
  infinitesimal photon mass {\tt MPHOTON}, in the sense that there is, up to
  irrelevant terms of order ${\cal O}(\epsilon)$, a one-to-one correspondence
  between the logarithmic term $\log {\tt MPHOTON}^2$ and the term
  $\Gamma(1+\epsilon)(4\pi \mu^2)^\epsilon/\epsilon$
without additional finite terms.}.

%

If regulator masses are used, the cut-off dependence of the loop amplitudes
is much less transparent than in the case of dimensional regularisation.  The
flag {\tt IRsubtraction}, introduced above, may be used to indicate that a IR
subtracted result is also provided.  If the functionality is not sufficient,
e.g.\ if QCD and EW corrections are treated by the same OLP, special
additional flags like {\tt EWIRregularisation} and {\tt EWIRsubtraction}
might be useful with similar keywords as the QCD counterparts but
supplemented by the ``EW'' prefix.

The OLP directly subtracts the collinear and soft divergences, for example
via the ``endpoint contributions'', quantified, for example, by the ${\bf
  I}(\epsilon)$~operator of Catani and Seymour, in the dipole subtraction for
QCD~\cite{Catani:1996vz,Phaf:2001gc,Catani:2002hc} or
QED~\cite{Dittmaier:1999mb,Dittmaier:2008md} corrections.  While this
introduces a dependence on the subtraction scheme, it renders the sum of the
virtual corrections and the endpoint of the real corrections independent of
any regularisation. As a consequence, the LO MC needs less information and
can, in particular, calculate the real corrections without any dependence on
the IR regularisation scheme.
%

\subsection{Electroweak renormalisation scheme}
The next point is the agreement on the renormalisation scheme which, in
essence, defines relations between electroweak couplings and parameters
beyond leading order. Electroweak radiative corrections are absorbed
differently in renormalised couplings in various schemes.
The default, supported by any OLP, should be the $\alpha(0)$
scheme for real~\cite{Belanger:2003sd, Aoki:1982ed, Bohm:1986rj,
Denner:1991kt} or complex~\cite{Denner:2005fg} on-shell masses and
electroweak couplings, where the weak mixing angle is defined via
$\cos\theta_W=M_W/M_Z$.  In this respect, it is essential that OLP
subprocesses have identifiable overall coupling factors like
$\alpha^N\,\as^M$ which is guaranteed by the correct use of the
flag {\tt PowerAlphas} together with a similar flag for $\alpha$,
{\tt PowerAlpha}.

In the simplest case, the MC calculates $| \mathcal{A}_{\LO}|^2$ with only
$\alpha(0)$ as EW coupling.  The OLP might use another value for $\alpha$
such that the two Born evaluations will not match. The ratio can be used to
infer an effective coupling $\aOLP$ which is defined by
\begin{equation}
\aOLP = \alpha(0) \left( \frac{\left| \mathcal{A}_{\LO}^{\rm
    OLP}\right|^2}{\left| \mathcal{A}_{\LO}^{\rm MC}\right|^2} \right)^{1/P}
\end{equation}
where $P$ is the value of the {\tt PowerAlpha} flag.  This defines the value
for $\alpha$ which is consistent with the OLP. A safer way would be to pass
$\aOLP$ as an additional parameter during the run-time phase.  In this way
the MC could still perform a consistency check between the Born cross
sections after adjusting the coupling parameters.


In general, e.g.\ if LO MC and OLP do not use the same masses or
$\sin^2\theta_W$, more information has to be exchanged.  The minimal
information needed by the MC includes the LO matrix element as computed by
the OLP and the quantities appearing in the IR modules.  Then the MC has all
elements to cancel the IR singularities properly.  The IR modules, basically,
depend on the electromagnetic coupling and the regularisation scheme. The LO
matrix element can, in principle, be taken from the OLP, LO(OLP). In
practice, however, the MC should be able to calculate the LO matrix element
independently. To this end, the MC needs all parameters entering the LO(OLP)
matrix element.

Therefore, the OLP specifies all parameters needed to calculate the LO matrix
element, $\mathcal{A}_{\LO}^{\rm OLP}$: couplings, masses, $\sin^2\theta_W$,
etc.  Masses for unstable particles are always understood as defined from
complex poles, i.e.\ including real and imaginary parts, $M_j^2 - i\,M_j
\Gamma_j$, but no running widths.

While the $\alpha(0)$ scheme should be the default for computing
$\mathcal{A}_{\LO}$, we should allow for other schemes to keep the interface
more general and to improve the accuracy of the predictions.  Various
electroweak renormalisation schemes are discussed in the literature.  The
$\alpha(M_Z^2)$ scheme
and the $\GF$ scheme
are defined, for instance, in Ref.~\cite{Dittmaier:2001ay}.  Similarly,
$\alpha(\mu^2)$ with some scale $\mu$
or other definitions such as $\overline{\mathrm{MS}}$
could be used.  Moreover, the OLP could also support other schemes where
different couplings are used in the same amplitude, like $\alpha(0)$ and the
Fermi constant $\GF$.
\begin{description}
\item[{\tt EWRenormalisationScheme}:] used schemes (discussed in the text)
  can be flagged by the keywords {\tt alpha0} (default), {\tt alphaMZ}, {\tt
    alphaGF}, {\tt alphaRUN}, {\tt alphaMSbar}, {\tt OLPdefined}.
\end{description}

If the electromagnetic coupling is defined by the OLP ({\tt OLPdefined}), the
OLP returns an effective $\aOLP$ ({\tt alphaOLP}) to be used in the LO matrix
element.  To give an example, consider the process $qq\to W\gamma$. Here the
effective $\alpha$ would result from $\aOLP^2 =\alpha(0)\,\alpha_{\GF}$,
where $\alpha_{\GF}$ is $\alpha$ derived from the Fermi coupling constant. In
some approaches, where the QED corrections are separated from the pure EW
part of the calculation, an $\aIR$ ({\tt alphaIR}) is introduced, which has
to be used in the IR modules multiplying LO matrix elements. In addition to
the effective $\aOLP$ that is used already in LO, $\aIR$ is the EW coupling
of the photonic correction only.  The coupling parameters $\aOLP$ and $\aIR$
should be returned by the OLP during the run-time phase, together with the
amplitude information. After ensuring cancellation of IR singularities, the
MC can use or not $\aOLP$ for real corrections, QCD contributions etc.

After this exchange of information the MC knows how to cancel IR
singularities and can proceed to calculate the remaining part of the real
corrections.  Clearly, the main issue is to transfer all the knowledge needed
by the MC for the calculation of this part of the NLO correction.  This
knowledge, in particular, includes the issues of complex masses, potentially
complex couplings etc.  Therefore, it is essential that the OLP is returning
not only the complex masses, but also explicitly the needed values of the
couplings, which can be either complex or real, according to the OLP choice.

\subsection{The run-time phase}

During the {\tt run-time} phase, the exchange of information proceeds as
follows:

\begin{itemize}

\item The MC calculates $\left| \mathcal{A}_{\LO}^{\rm MC}\right|^2$ for a
  given process, according to its internal conventions (e.g.\ $\alpha(0)$ as
  EW coupling).
\item The OLP computes $\left| \mathcal{A}_{\LO}^{\rm OLP}\right|^2$ with an
  internal choice of EW couplings and parameters and the interference term
  $2\,{\rm Re}\bigl[ \mathcal{A}_{\LO}^{\rm
      OLP}\bigr]^{\dagger}\,\mathcal{A}_{\NLO}^{\rm OLP}$.  It returns these
  values according to the definitions of Tab.~\ref{inoutcontrol}, i.e.\ the
  EW correction split up into coefficients of the poles in $\epsilon$.
\item The OLP returns an effective $\aOLP$ ({\tt alphaOLP}) and, if required,
  an $\aIR$ ({\tt alphaIR}). These values may change event by event, if
  variable scales are used for their definition.
\item The MC uses $\aOLP$ for the LO matrix element and
  $\aIR$ for the IR modules. Alternatively, it can determine
  $\aOLP$ from the evaluation of the Born amplitudes by the OLP
  and the MC.
\end{itemize}
Note that, in the case of mass regularisation, the coefficient of the double
pole in $\epsilon$ is zero.

As the correct use of coupling parameters in EW computations is a complicated
task, the MC should have the functionality to evaluate all necessary
amplitudes in a ``coupling stripped'' mode, such that the couplings can be
imported by the OLP during run-time.  Sometimes it might be useful for the MC
to use the Born term which is returned by the OLP.  We remind that the
command {\tt OLP\_EvalSubProcess} can be adapted to pass additional
information between the MC and the OLP.

It is clear from the discussion that a standardised interface between a MC
and an OLP is a non-trivial task for electroweak processes, because much more
cross talk between the computer programs is necessary. We would like to
motivate research groups, which are active in this field, to provide public
examples of their individual solutions such that the presented framework can
be further improved in the future.

\section{Summary and outlook}
This paper summarises the discussion that the ``next-to-leading order
multi-leg'' group had during the workshop on ``Physics at TeV colliders'' at
Les Houches 2009 on how to interface tree-level Monte Carlo event generators
with computer codes that provide one-loop information.  Due to the fast
progress in the field over the last few years, such a discussion was
desirable, as it became more and more evident that synergies between the MC
and the loop calculator community can speed-up the production rate of higher
order computations considerably.  One-loop precision for a large number of
QCD and electroweak observables is already available, and some of these
computations already use process-specific interfaces.  Automated approaches
for one-loop amplitude computations are pursued by several groups and recent
progress shows that efficient and numerically reliable implementations exist.
A standardised interface will not only dispense the loop-calculator from
doing the rather involved tree-level computations: it eventually will allow
to efficiently import one-loop information into state of the art Monte Carlo
tools.  Due to the modular set-up, exchange of code from various providers
will speed-up debugging and validation of computations considerably.

Following our extensive discussion, we have presented, in this paper, the
first proposal for a standardised interface between leading order Monte Carlo
tools and one-loop programs.  Several program authors are committed to
implement the proposed interface at the moment. The first experiences are
very promising and will be documented in the Les Houches proceedings.

The current proposal must not be regarded as the endpoint of a
standardisation discussion. As nothing is carved in stone yet, modifications
of various details are likely to improve the interface in the future, e.g.\
in/output formats might be specified differently.  It is planned to make some
program fragments public on the Les Houches web page:
\begin{verbatim}
http://www.lpthe.jussieu.fr/LesHouches09Wiki/index.php/Draft
\end{verbatim}

A natural extension of the proposed interface is to include information which
is relevant for adding a parton shower on top of the proposed NLO fixed order
accord. This rather concerns the treatment of the IR subtraction and is thus
orthogonal to the passing of one-loop information.  Discussion on this issue
has already started during the 2009 Les Houches workshop.

Finally we would like to invite the community to give us feedback on this
initiative. We are committed to incorporate good proposals and constructive
comments in later versions. We believe that this initiative will, in the end,
lead to a more efficient use of resources and hopefully will help to describe
a wide range of Tevatron and LHC data with Monte Carlo tools beyond the
leading order level.

\bibliography{NLOLHA_v2}

\end{document}